# Maximizing Savonius Turbine Performance using Kriging Surrogate Model and Grey Wolf-Driven Cylindrical Deflector Optimization

Paras Singh[1*], Vishal Jaiswal[1], Subhrajit Roy[1], and Raj Kumar Singh[1]

[1]Department of Mechanical Engineering, Delhi Technological University, Delhi-110042, India
*parass2802@gmail.com; viztech2002@gmail.com; roysubhrajit03@gmail.com; rajkumarsingh@dce.ac.in

**ABSTRACT**

With the growing power demand and the imperative for renewable energy sources, wind power stands as a vital component of the energy transition. To optimize energy production, researchers have focused on design optimization of Savonius-type vertical axis wind turbines (VAWTs). The current study utilizes Unsteady Reynolds-Averaged Navier Stokes (URANS) simulations using the sliding mesh technique to obtain flow field data and power coefficients. A Kriging Surrogate model is trained on the numerical data of randomly initialized data points to construct a response surface model. The Grey Wolf Optimization (GWO) algorithm is then utilized to achieve the global maxima on this surface, using the turbine's power coefficient as the objective function. A comparative analysis is carried out between simulation and experimental data from prior studies to validate the accuracy of the numerical model. The optimized turbine-deflector configuration shows an improvement of 34.24% in power coefficient. Additionally, the GWO algorithm's effectiveness is compared to Particle Swarm Optimization (PSO) and is found to be better in most cases, converging towards the global maxima faster. This study explores a relatively unexplored realm of metaheuristic optimization of wind turbines using deflectors, for efficient energy harvesting, presenting promising prospects for enhancing renewable sources.

## 1. INTRODUCTION

The pressing need to address environmental challenges and combat climate change has driven a significant shift towards green technologies. With a growing emphasis on finding sustainable and environmentally friendly energy sources, there is a rising interest in adopting these solutions worldwide. One particularly noteworthy and practical renewable energy source is wind power [1]. Due to its potential for use in areas with low wind speeds, vertical axis wind turbines (VAWTs) have drawn more interest than their horizontal axis equivalents. Even still, VAWTs' poor wind energy conversion efficiency is still a problem [2]. At present, researchers are dedicated to improving the performance of VAWTs, with a special focus on the Savonius turbine. This turbine's distinctive design enables it to operate efficiently even in areas with low wind resources, eliminating the need for a self-starting mechanism.

## 2. LITERATURE REVIEW AND OBJECTIVE

Sigurd Johannes invented the Savonius turbine, which is a drag-based turbine with S-shaped blades rotating around an axis. Researchers are actively exploring methods to improve its efficiency, which is evaluated by $C_P$ (coefficient of power) and $C_M$ (coefficient of moment).

$$C_M = \frac{\text{Moment}_{turbine}}{0.5\rho AV^2 R} \quad (1)$$

$$C_P = C_M \times TSR \quad (2)$$

Here A is the frontal area of the turbine, V is the freestream velocity, R is the radius of the turbine, $\rho$ is the density of the medium, and TSR is the ratio of the tip speed of the turbine and the freestream velocity. The reverse torque created by the returning rotor is one of the main problems found in the Savonius turbine, which affects its overall efficiency. To address this, researchers propose blade modifications by altering shape or profile to capture energy more effectively. Additionally, they are studying the use of deflectors to guide wind flow and enhance performance. Both experimental and numerical methods are employed for the optimization studies. Experiments with physical models offer insights into fluid dynamics, while CFD simulations predict flow patterns and aid in design improvements. Recent advances include the utilization of nature-based algorithms like PSO and GA, along with machine learning models for further optimization. Kassem et al. [3] improved the end plate designs of the Savonius VAWT and conducted CFD simulations for different wind speeds. When opposed to possessing no end plates, installing both upper and lower end plates led to a 35 percent rise in rotor power. Zemamou et al. [4] optimized the blade design using Bezier curves and CFD simulations, achieving an impressive 29% improvement in power coefficient compared to the traditional VAHT. Xia et al. [5] optimized the blade form using nature-based algorithms and surrogate models, resulting in an average 7% improvement in torque coefficient compared to the old design. He et al. [6] used Evolutionary-based Genetic Algorithms (GAs) with CFD simulations to optimize the blade form and deflector location of the Savonius VAWT. When compared to semi-circular blades, the optimal blades improved the time-averaged coefficient of power by 34%. Furthermore, as compared to the baseline turbine configuration, the improved deflector resulted in an incredible 95% improvement in the time-averaged coefficient of power. Golecha and Prabhu [7] conducted experiments on the impact of a rectangular deflector's position near the Savonius Water turbine. Placing the deflector at an optimized position resulted in a substantial 50% increase in the coefficient of power, demonstrating the importance of precise deflector placement in enhancing the turbine's performance. However, solid deflectors come with a notable drawback: they generate high turbulence and wake regions behind them, which significantly impacts flow dynamics, especially in the vicinity of the returning blade. This leads to negative torque production and a dramatic reduction in the VAWT's power and torque. Flat plate deflectors can create large vortices behind them, causing flow instability and affecting the returning blade. Moreover, the interaction between the wind turbine and deflector causes cyclic load fluctuations, leading to increased system fatigue. To address these challenges, researchers have been focusing on circular deflectors as a potential solution for Savonius VAWTs. The use of circular deflectors aims to break down wake zones, reduce downstream vortices, and improve overall turbine performance. Yuwono et al. [9] found that using circular deflectors upstream of the returning blade resulted in a 12.2% improvement in $C_p$ at a TSR of 0.65 when compared with a standard Savonius turbine. Setiawan et al. [10] proposed attaching a circular cylinder to the revolving rotor of the Savonius turbine to create a nozzle-like effect that propelled the rotating blade and increased positive torque. Their numerical investigation found that the coefficient of power had gone up by approximately 17.3%. Fatahian et al. [8] investigated the impact of positioning a rotating circular deflector upstream of the returning blade in a Savonius VAWT. They observed that the turbine-deflector system's performance is sensitive to specific parameters, namely $D_c$ (deflector's diameter), $L_x$ (horizontal distance of deflector from turbine's shaft), and $L_y$ (vertical distance of deflector from turbine's shaft). The purpose of this paper is to optimize these critical parameters in the Savonius VAWT for a stationary circular cylindrical deflector located upstream of the returning blade. This work aims to produce an efficient and improved performance for the turbine-deflector combination by combining the kriging surrogate model and Grey Wolf optimization with CFD.

## 3. COMPUTATIONAL MODELLING
### 3.1 Deflector Layout and Parameterization

For the current optimization study, a Vertical Axis Savonius turbine having a cylindrical deflector placed upstream of the returning blade was used. Figure 1 illustrates the turbine's schematic representation with the deflector strategically positioned. The turbine's specifications, including $D_b$=0.5m, and D=0.909m, were based on Yuwono et al.'s experiments [9], where D and $D_b$ are the diameters of the turbine and the individual blades respectively. The experiments revealed that incorporating the deflector significantly improves the turbine performance, influenced mainly by three parameters: the cylinder's diameter ($D_c$), its horizontal distance from the origin ($L_x$), and its vertical distance from the origin ($L_y$). Previous experimental and numerical studies [9, 10] have shown that precise adjustments of these parameters can substantially enhance the turbine's performance. To achieve optimal performance and efficiency for the turbine-deflector system, the parameters should be within specific ranges determined from previous investigations [8]: $L_x/D$ between 1 and 2, $L_y/D$ between 0.3 and 1, and $D_c/D$ between 0.25 and 1. Since the performance of the system is sensitive to its design parameters, adhering to these defined ranges is crucial for maximizing the turbine's output.

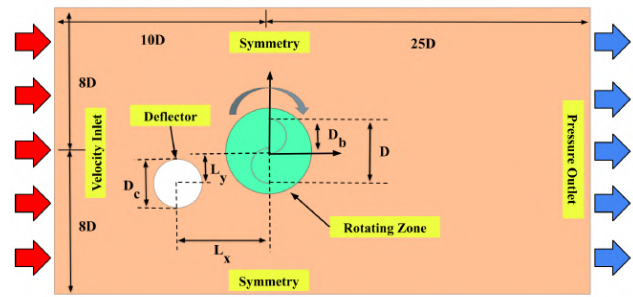

**Figure 1: Schematic diagram of the domain**

### 3.2 Computational Domain

To conduct the CFD simulations, a rectangular domain was utilized, centered on the turbine's shaft. It was divided into two subdomains: one for the rotating inner part (diameter $D_r$=1.2m) with the turbine blades and another for the larger stationary outer region. This separation enabled a focused analysis of fluid flow interactions in the Savonius turbine. For mesh generation, the computational domain was split into two sections: the inner zone, containing the deflector and turbine rotors, was meshed with unstructured tri elements, while the outer part used structured quad elements. This approach was chosen to avoid poor-quality cells and high aspect ratio elements near the turbine walls and the deflector. The domain partition also allowed for the adoption of a parametric framework, streamlining the simulations. For each new design point with distinct values of $D_c$, $L_x$, and $L_y$, the procedure involved sequential updates of geometry, remeshing, and solution processes. This meshing technique allowed the mesh generator to focus solely on updating the mesh within the inner domain while keeping the outer domain mesh unchanged, significantly improving computational efficiency. While generating the mesh, a crucial consideration was maintaining a $y^+$ value below 1 to accurately capture boundary layer effects. Around the turbine and the deflector, 32 inflation layers were wrapped. The first layer had a height of $5\times10^{-3}$ mm, and the following layer heights were determined using a constant growth rate of 1.14. Figure 2 illustrates the cell distribution within the flow domain of the turbine-deflector system. The refined grid near the turbine and deflector walls aimed to accurately resolve the flow physics in regions of high turbulence.

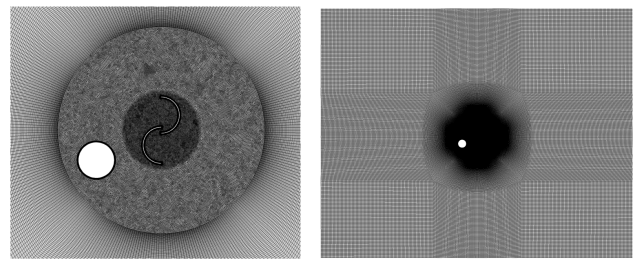

**Figure 2: Computational Grid**

## 3.3 Numerical Model and Boundary Conditions

Unsteady RANS simulations were performed using Ansys Fluent 22.1. The selection of a suitable turbulence closure model greatly impacts simulation accuracy and stability. Previous numerical investigations [2, 3, 4, 6, 7, 8] have consistently shown that Menter's k-ω Shear Stress Transport (SST) turbulence model [13] is well-suited for representing VAWTs behavior accurately. The model utilizes a blending function to switch between the Wilcox k-ω model [14] (close to a wall) and the standard k-ϵ model [15] (far field region). This blending leverages the robust near-wall formulation of the Wilcox k-ω model while mitigating its sensitivity to freestream conditions, while also utilizing the k-ϵ model's advantage of freestream independence. A coupled approach was used for pressure-velocity coupling, in which the continuity and momentum equations are solved simultaneously instead of the predictor-corrector approach. This reduces the convergence time and enhances the solution stability and reliability. For spatially discretizing the transport quantities, a second-order upwind technique was utilized, combined with a least squares cell-based approach for gradient discretization. A freestream velocity of 7 m/s was assigned at the inlet, while the outlet condition was set as a pressure outlet. The deflector and turbine blade walls were set with a no-slip condition. Constant fluid density (1.225 kg/m$^3$), dynamic viscosity (1.7894 × 10$^{-5}$ kg/m.s), and operating pressure (101325 Pa) were taken throughout all the simulations. The data from these simulations were used to train the surrogate model for the subsequent optimization study.

## 3.4 Validation Studies

In this study, a grid independence analysis was conducted to ensure accurate results in numerical simulations, considering various grid resolutions. While in general, a fine grid provides more accurate results, it comes with a higher computational cost. Therefore, the objective was to determine an ideal mesh size that strikes a balance between accuracy and simulation time. Figure 3 (a) represents the variation of the moment with the rotation angle of the turbine for three different mesh resolutions (66000, 123000, 230000). The results demonstrate that both medium and fine meshes yield similar results, and further refinement would have no impact on the solution. Therefore, a mesh size of 123000 elements can effectively predict the turbine's flow physics.

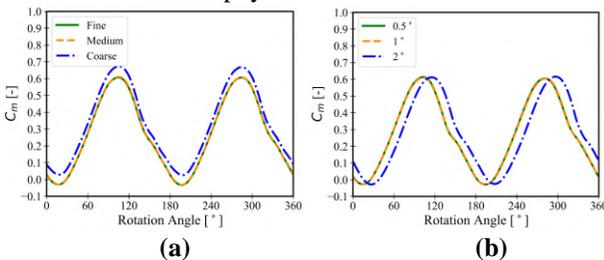

**Figure 3: Grid Independence Study (b) Time Step Independence Study**

Three different time steps: 0.5°, 1°, and 2° turbine rotation per time step were utilized for analyzing the effect of time step size on the solution. Figure 3(b) represents the variation of $C_p$ with the turbine's rotation angle for different time step sizes. It is evident that decreasing the time step below 1° per time step will not cause any significant effect on the solution, and thus, this time step is used for the subsequent simulations. Sheldahl et al. [16] experiments and Fatahian et al. [8] numerical investigations were utilized to validate the accuracy of the numerical model. From Figures 4(a) and 4(b), it can be observed that the results of the current study closely match the experimental results, with very small discrepancies. Thus, it can be inferred that the numerical model used in the present study is independent of time step and mesh sizes, and is in close agreement with the experimental results.

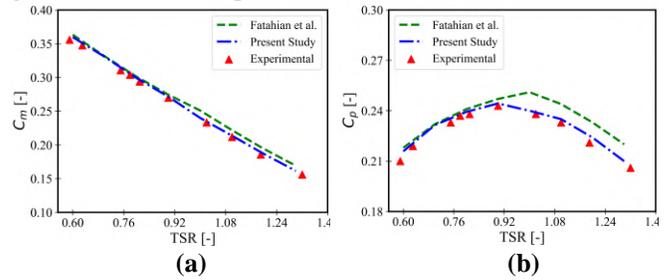

**Figure 4: Variation of (a) $C_M$ and (b) $C_P$ with TSR of the turbine**

## 4. OPTIMIZATION FRAMEWORK

### 4.1 Sample Acquisition

The acquisition of samples is a vital component of the optimization process since it forms the basis for the subsequent analysis. The primary objective is to gather relevant data points accurately representing the system's behavior. The study aims at enhancing the performance of Savonius wind turbines by placing a circular cylindrical deflector upstream of the blades. To acquire samples, numerical simulations are conducted to capture the turbine's behavior under various operating conditions and design parameters such that points are spread evenly to represent the entire design space.

### 4.2 Surrogate Modeling

In the context of optimization, conducting a large number of computer simulations can be prohibitively expensive due to the need for substantial computational resources. This is where surrogate models play a vital role. Surrogate models act as efficient approximations of the actual system behavior, allowing us to streamline the optimization process. The surrogate model technology uses regression methods to infer the response value of unknown design points based on known sample points. The primary objective is to construct an approximate model that achieves adequate prediction and fitting accuracy [5]. Previous studies [5, 11] have employed the Radial Basis Function surrogate model to predict significant nonlinear interactions between variables. The present study utilizes the Kriging surrogate model. Kriging is particularly well-suited for applications involving complex and nonlinear behaviors, making it an ideal choice for optimizing turbine performance.

### 4.3 Grey Wolf Optimization

The Grey Wolf Optimization (GWO) is a nature-inspired optimization algorithm. It takes inspiration from the hunting behaviors and social hierarchical structure of Grey wolves. It was proposed by Seyedali Mirjalili et al. [12] in 2014. The algorithm imitates the natural hunting behavior of Grey wolves and adopts a leadership hierarchy similar to a wolf pack. The hierarchical levels consist of alpha, beta, delta, and omega wolves. This hierarchical structure serves as the foundation for the search mechanism in the algorithm. To find the optimal parameters for the circular cylinder deflector, we employ the Grey Wolf Optimization (GWO) algorithm.

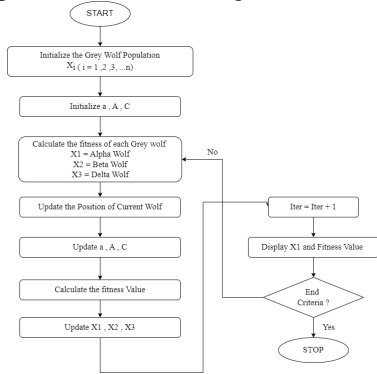

**Figure 5: Grey Wolf Optimization Algorithm**

During the optimization procedure, GWO iteratively searches for the global optimum in the design space by iteratively updating the positions of alpha, beta, delta, and omega wolves. These positions correspond to the optimal solutions for the problem. GWO uses the Kriging surrogate model to evaluate the objective function efficiently, avoiding the need for computationally expensive evaluations.

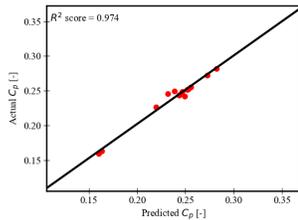

**Figure 6: Validation of the Kriging Model**

### 5. RESULTS AND DISCUSSION

#### 5.1 Kriging Surrogate Model

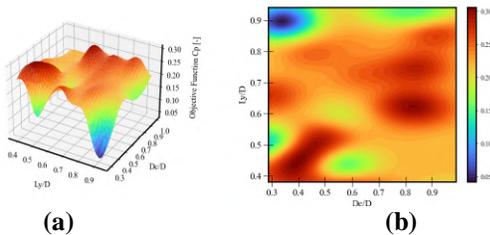

(a)  (b)

**Figure 7: Response surface generated using the Kriging model**

For the present study, a total of 45 sample points were generated, and CFD simulations were specifically conducted at TSR (Tip Speed Ratio) 0.9, based on the findings of a previous study by Fatahian et al. [8] that indicated peak values at this particular point. Among the 45 points, 30 were chosen for the training set, while the remaining points were utilized to validate the surrogate model. The performance of the kriging model was assessed, resulting in an $R^2$ score of 0.974, as depicted in Figure 6.

Three scenarios were taken into consideration to examine how design parameters affected the objective function. To create a response surface, one of the three factors ($D_c/D$, $L_x/D$, & $L_y/D$) was kept constant in each scenario. The surrogate model generated response surfaces for each case, revealing that the parameter $L_x/D$ had minimal impact on the design. However, Figure 7 clearly illustrates a strong non-linear correlation between the objective function and the design parameters $D_c/D$ and $L_y/D$. The response surface indicated that optimal solutions were obtained for values of $D_c/D$ between 0.35 and 0.45, and $L_y/D$ between 0.4 and 0.5. These findings offer valuable insights for optimizing the design parameters of circular cylinders, as they identify the ranges of $L_y/D$ and $D_c/D$ that significantly enhance the $C_p$ of the Savonius turbine.

#### 5.2 Grey Wolf Probability Parameter

By training a surrogate model with CFD data using kriging model imported from the open-source SMT Python package [17] and utilizing an in-house developed code for Grey Wolf Optimization (GWO) algorithm, the objective function was maximized. The GWO algorithm relies on a carefully chosen probability parameter ($P_a$) to ensure its effectiveness. Figure 8(a) illustrates the influence of the probability parameter on the algorithm's performance. The convergence of the algorithm was analyzed for $P_a$ values of 0.1, 0.25, 0.4, 0.5, 0.75, and 0.8. From the results shown in Figure 8, it is evident that GWO performs optimally when $P_a$ is set to 0.4. This choice enables the algorithm to converge much faster and yield the most optimal solution compared to other values of $P_a$.

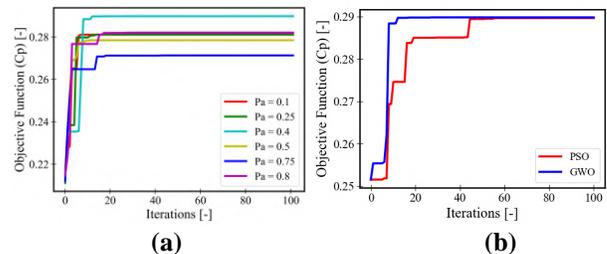

(a)  (b)

**Figure 8: (a) Performance of GWO algorithm for different values of probability parameter (b) Performance of GWO algorithm against PSO algorithm.**

#### 5.3 Performance of GWO against PSO

The outcomes of the GWO algorithm were compared with those of the Particle Swarm Optimization (PSO) method, another well-known algorithm. The population size was set to 25 in both situations. Particle Swarm Optimization (PSO) is an optimization algorithm that draws its inspiration from flocks of birds and other social phenomena. It performs iterative search space exploration using a population of particles. To achieve the optimum outcome, each particle modifies its position based on local and global best-known solutions. Due to its ease of use and success in tackling a variety of optimization issues, PSO is widely used. As depicted in Figure 8(b), it was observed that both the GWO and PSO algorithms nearly reached the same optimal solution, but GWO had a slight advantage over PSO. However, there was a notable difference in convergence behavior between the two. GWO appeared to converge after only 15 iterations, while PSO required more than 50 iterations to reach the optimal solution.

### 5.4 Optimum Turbine Parameters-

The results of the optimization study are presented in Figure 9. The graph compares the values of $C_p$ variation with TSR for three cases: the baseline turbine without a deflector, the configuration with initial design parameters for the optimization algorithm ($L_x/D$= 1.08, $L_y/D$= 0.38, $D_c/D$= 0.90), and the optimized parameters ($L_x/D$= 1.21, $L_y/D$= 0.45, $D_c/D$= 0.38). It is evident from the graph that the optimized configuration outperforms the other two cases for all TSR values, with a significant improvement in the value of $C_p$ compared to the baseline case. At a TSR of 0.9, the optimized configuration shows a remarkable 34.24% increase in $C_p$ compared to the baseline turbine.

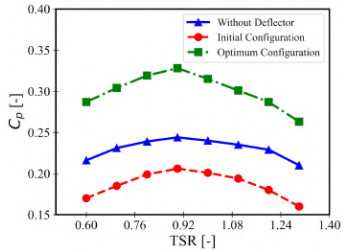

**Figure 9: $C_p$ vs TSR distribution for different turbine configurations**

### 5.5 Flow Structure

For the analysis of flow physics and performance enhancement using deflectors, pressure, velocity, and vorticity contours are presented at a rotational angle of 300 degrees and a TSR of 0.9, which corresponds to the peak of $C_p$. In Figure 10, the pressure distribution is shown for both the baseline configuration (without a deflector) and the configuration with the optimized cylindrical deflector. In the absence of a deflector, the incoming airflow from the left side strikes both the advancing and returning blades, resulting in a net torque that drives the turbine in a clockwise direction. This occurs because the flow impinging on the turbine blades causes the advancing blade to rotate clockwise and the returning blade to rotate anti-clockwise. The net drag force acting on the advancing blade is higher compared to the returning blade due to its blade shape, leading to the overall clockwise rotation of the turbine. However, when a cylindrical deflector is positioned ahead of the returning blade, the airflow gets split as it comes into contact with the deflector, due to which the drag acting on the returning blade is reduced. This effect is evident from the pressure contours presented in Figure 10; for the case without a deflector, the pressure is larger at the convex portion of the returning blade. Additionally, the pressure at the downstream end of the turbine is lower for the advancing blade and higher for the returning blade when a deflector is used. Consequently, this increases the net driving torque on the turbine, resulting in a higher generated power output.

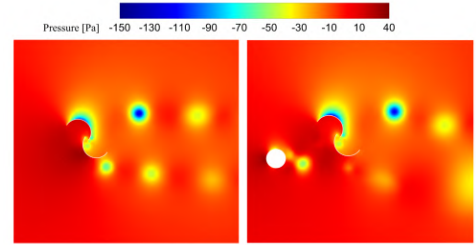

**Figure 10: Pressure distribution for the baseline and optimized turbine-deflector configuration at TSR=0.9**

The turbine rotor's rotation creates two separate regions, the high-velocity area and the wake area, which are shown in Figure 11 in the form of velocity contours around the system. The baseline case and the turbine-deflector system are found to be significantly different. Relative to the baseline design without a deflector, the highest velocity near the leading edge of the advancing blade is greater in the turbine-deflector system. Due to increased pressure drag and consequently higher net driving torque on the turbine, this causes a decrease of pressure at the downstream end of the driving blade. Additionally, the stagnant region upstream of the turbine is less prominent when the cylindrical deflector is introduced. This can be attributed to the vortex-shedding effect from the cylindrical deflector, which imparts momentum to the stagnant region, causing it to be less significant compared to the baseline case.

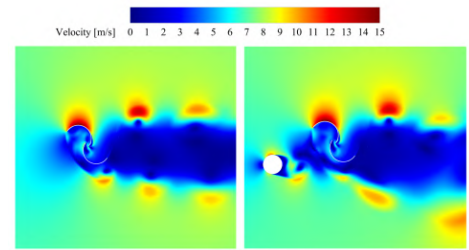

**Figure 11: Velocity distribution for the baseline and optimized turbine-deflector configuration at TSR=0.9**

Figure 12 presents the vorticity contours, allowing visualization of the Karman vortex street behind the cylindrical deflector and the vortex shedding from the Savonius rotor. The presence of the cylindrical deflector influences the flow pattern, directing it towards the advancing blade and below the returning blade. As a result, the deflector acts as a shield for the returning blade, providing additional evidence of the increased torque effect.

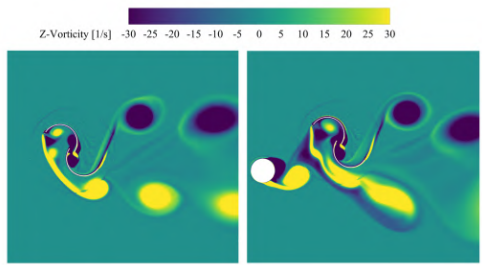

**Figure 12: Vorticity contours for the baseline and optimized turbine-deflector configuration at TSR=0.9**

## 6. CONCLUSIONS

The paper provides a comprehensive optimization framework intended to improve the performance of a Savonius wind turbine with a cylindrical deflector. To achieve this, a multifaceted approach involving surrogate modeling, computational fluid dynamics (CFD) simulations, and metaheuristic optimization algorithms has been employed. For training the Kriging surrogate model, 45 design points were generated through random initialization, and URANS simulations were run on them at a TSR of 0.9. While comparing optimization algorithms, the Grey Wolf algorithm outperformed the Particle Swarm Algorithm. It efficiently converged to optimal design parameters for the cylindrical deflector, leading to its selection for parameter optimization. The optimized cylindrical deflector had a $D_c/D$ of 0.38, $L_y/D$ of 0.45, and $L_x/D$ of 1.21. Compared to the standard Savonius turbine, the maximum $C_p$ increased from 0.244 to 0.328. This substantial improvement serves as a testament to the efficacy of the proposed optimization framework in carrying out optimization for such complex design spaces.

However, there are some limitations to this framework. The use of URANS simulations, instead of high-fidelity models like DES and LES, limits the accuracy of the results. It is also important to acknowledge that surrogate models introduce an element of approximation, which, in certain cases, can result in lower accuracy. While the surrogate model used in the present study served as a valuable tool to expedite the optimization process, a nuanced trade-off exists between computational efficiency and the precise representation of the underlying physics. Despite these limitations, the findings of this study remain noteworthy. The achieved results not only advance the current understanding of wind turbine design but also hold promise for practical applications in other areas related to renewable energy systems. The methodologies embedded within the framework signify that the relevance of the framework transcends the immediate scope of this study.

## NOMENCLATURE

| | | |
|---|---|---|
| $D$ | Turbine Diameter | [m] |
| $D_c$ | Diameter of Deflector | [m] |
| $L_x$ | Horizontal Distance of Deflector from the central point of the turbine | [m] |
| $L_y$ | Vertical Distance of Deflector from the central point of the turbine | [m] |
| $C_P$ | Power coefficient | -- |
| $\omega$ | Angular velocity of the turbine | [rad/s] |
| $C_M$ | Moment coefficient | -- |
| TSR | Tip Speed Ratio | -- |

## REFERENCES

[1] G.M. Shafiullah et al., Prospects of Renewable Energy – A feasibility study in the Australian context, Renewable Energy, 39(1), 2012, pp. 183–197.
[2] P.K. Talukdar et al., Parametric analysis of model Savonius hydrokinetic turbines through experimental and computational investigations, Energy Conversion and Management, 158, 2018, pp. 36–49.
[3] Y. Kassem et al., Performance investigation of Savonius Turbine with New Blade Shape : Experimental and Numerical study, International Journal of Applied Engineering Research, 13(10), 2018, pp. 8546–8560.
[4] M. Zemamou et al., A novel blade design for Savonius wind turbine based on polynomial bezier curves for aerodynamic performance enhancement, International Journal of Green Energy, 17(11), 2020, pp. 652–665.
[5] H. Xia et al., Blade shape optimization of savonius wind turbine using radial based function model and Marine Predator algorithm, Energy Reports, 8, 2022, pp. 12366–12378.
[6] D.Q. He et al., Performance-based optimizations on Savonius-type vertical-axis wind turbines using genetic algorithm, Energy Procedia, 158, 2019, pp. 643–648.
[7] K. Golecha, T.I. Eldho, and S.V. Prabhu, Influence of the deflector plate on the performance of modified savonius water turbine, Applied Energy, 88(9), 2011, pp. 3207–3217.
[8] E. Fatahian et al., An innovative deflector system for drag-type Savonius turbine using a rotating cylinder for performance improvement, Energy Conversion and Management, 257, 2022, p. 115453.
[9] T. Yuwono et al., Improving the performance of Savonius wind turbine by installation of a circular cylinder upstream of returning turbine blade', Alexandria Engineering Journal, 59(6), 2022, pp. 4923–4932.
[10] P.A. Setiawan, T. Yuwono, and W.A. Widodo, Numerical simulation on improvement of a Savonius vertical axis water turbine performance to advancing blade side with a circular cylinder diameter variations, IOP Conference Series: Earth and Environmental Science, 200, 2018, p. 012029.
[11] A. Tyagi, P. Singh, A. Rao, G. Kumar, and R.K. Singh, A Novel Framework for Optimizing Gurney Flaps using RBF Neural Network and Cuckoo Search Algorithm, arXiv preprint, 2023, arXiv:2307.13612.
[12] S. Mirjalili, S.M Mirjalili and A. Lewis, Grey Wolf Optimizer, Advances in Engineering Software, 69, 2014, pp. 46-61.
[13] F.R. Menter, Two-equation eddy-viscosity turbulence models for engineering applications, AIAA Journal, 32(8), 1994, pp. 1598–1605.
[14] D.C. Wilcox, Formulation of the k-ω turbulence model revisited', AIAA Journal, 46(11), 2018, pp. 2823–2838.
[15] B.E. Launder and D.B. Spalding, The numerical computation of turbulent flows, Computer Methods Applied Mechanics and Engineering, 1974, pp. 269-289.
[16] R.E. Sheldahl, B.F. Blackwell, and L.V. Feltz, Wind Tunnel performance data for two- and three-bucket Savonius rotors, Journal of Energy, 2(3), 1978, pp. 160–164.
[17] P. Saves et al., SMT 2.0: A Surrogate Modeling Toolbox with a focus on Hierarchical and Mixed Variables Gaussian Processes, arXiv preprint, 2023, arXiv.2305.13998